\documentclass[review]{elsarticle}
\usepackage{hyperref,graphicx,amsmath,amssymb,color,comment,amscd,amsfonts,undertilde,wrapfig,gensymb,lipsum}
\def\>{\rangle}
\def\<{\langle}
\def\n{\nonumber}

\begin{document}
\begin{frontmatter}
\title{Endo-reversible heat engines coupled to finite thermal reservoirs: A rigorous treatment}
\author[mymainaddress]{Ilki Kim\corref{mycorrespondingauthor}}
\cortext[mycorrespondingauthor]{Corresponding author}
\ead{hannibal.ikim@gmail.com}

\author[mysecondaryaddress]{Hui Wan}

\author[mysecondaryaddress]{Soumya S. Patnaik}

\address[mymainaddress]{Center for Energy Research and Technology, North Carolina
A$\&$T State University, Greensboro, NC 27411, U.S.A.}
\address[mysecondaryaddress]{Aerospace Systems
Directorate, Air Force Research Laboratory, Dayton, OH 45433,
U.S.A.}
\begin{abstract}
We consider two specific thermodynamic cycles of engine operating in
a finite time coupled to two thermal reservoirs with a {\em finite}
heat capacity: The Carnot-type cycle and the Lorenz-type cycle. By
means of the endo-reversible thermodynamics, we then discuss the
power output of engine and its optimization. In doing so, we treat
the temporal duration of a single cycle rigorously, i.e., without
neglecting the duration of its adiabatic parts. Then we find that
the maximally obtainable power output ${\mathcal P}_m$ and the
engine efficiency $\eta_m$ at the point of ${\mathcal P}_m$
explicitly depend on the heat conductance and the compression ratio.
From this, it is immediate to observe that the well-known results
available in many references, in particular the
(compression-ratio-independent) Curzon-Ahlborn-Novikov expressions
such as $\eta_m \to \eta_{\mbox{\tiny CAN}} = 1 -
(T_{\scriptscriptstyle L}/T_{\scriptscriptstyle H})^{1/2}$ with the
temperatures $(T_{\scriptscriptstyle H}, T_{\scriptscriptstyle L})$
of hot and cold reservoirs only, can be recovered, but,
significantly enough, in the limit of a {\em vanishingly} small heat
conductance and an {\em infinitely} large compression ratio only.
Our result implies that the endo-reversible model of a thermal
machine operating in a finite time and so producing a finite power
output with the Curzon-Ahlborn-Novikov results should be limited in
its own validity regime.
\end{abstract}

\begin{keyword}
%\texttt{elsarticle.cls}\sep \LaTeX\sep Elsevier \sep template
endo-reversible thermodynamics \sep finite thermal reservoir \sep
Carnot cycle \sep Lorenz cycle
%\MSC[2010] 00-01\sep 99-00
\end{keyword}
\end{frontmatter}

%\linenumbers

\section{Introduction}\label{sec:introduction}
The celebrated Carnot cycle is, as well-known, the very
thermodynamic cycle which reveals the maximally obtainable
efficiency ($\eta_{\mbox{\tiny C}}$) of all conceivable
work-producing engines of thermodynamics, being explicitly given by
\begin{equation}\label{eq:carnot-cycle_1}
    \eta_{\mbox{\tiny C}} = 1 - \frac{T_{\scriptscriptstyle L}}{T_{\scriptscriptstyle H}}\,,
\end{equation}
where the symbols $T_{\scriptscriptstyle H}$ and
$T_{\scriptscriptstyle L}$ denote the absolute temperatures of hot
and cold reservoirs, respectively \cite{CAR24,CAL85}. This cycle
consists of four different thermodynamic processes, being two
isothermal and two adiabatic ones, each of which operates
reversibly, or infinitesimally slowly, thus leading the temporal
duration $t_{cy}$ of a single cycle to going to an {\em infinity}.
Due to this non-realistic setup, the Carnot cycle cannot produce a
finite power output ${\mathcal P} = W/t_{cy}$, though, where $W$ is
the net work output produced during a single cycle.

The second non-realistic aspect of the Carnot cycle with Eq.
(\ref{eq:carnot-cycle_1}) is the fact that the heat source and heat
sink are ideally modeled by two {\em infinite} thermal reservoirs
with the fixed temperatures ($T_{\scriptscriptstyle H}$ and
$T_{\scriptscriptstyle L}$), or a heating fluid and a cooling fluid
with an {\em infinitely} large heat capacity. This means that the
heat source and sink for a generic scenario should be spatially
enormous. For real-life thermal machines with practical
applications, however, this cannot be the case; e.g., a typical
power plant comes into play by exchanging heat in its compartments
of evaporation and condensation confined within a finite spatial
size (e.g., \cite{BEJ88,BEJ98,BEJ01}).

To circumvent the two aforementioned non-realistic factors of
drawback imposed upon the Carnot cycle in original form, researchers
have been paying considerable attention to the modified Carnot-type
cycle as a simple theoretical model of the non-ideal engine which
runs in a {\em finite} time as well as couples to the {\em finite}
heat source and sink. To this end, finite-time thermodynamics has
been under consideration, which has been applied successfully to
various irreversible processes \cite{SAL90,SIE94,SPA98,BER99}, such
as those of heat exchange in generic models of thermodynamic cycles
with a finite period $t_{cy}$, thus enabling us to determine the
performance bounds and optimal paths of such cycles for the sake of
optimized energy management.

One of the most widespread methods in form of the finite-time
thermodynamics is the {\em endo-reversible} approach in which the
working fluid of an engine exchanges heat irreversibly with the
heating and cooling fluids while it is operated {\em internally
reversibly} by an external driving in the Carnot limit
\cite{HOF97,HOF08}. Then it is well-known that within two infinite
thermal reservoirs, the cycle efficiency $\eta_m$ at the point of
the maximally obtainable power output ${\mathcal P}_m$ is given by
the so-called Curzon-Ahlborn-Novikov efficiency $\eta_{\mbox{\tiny
CAN}} = 1 - (T_{\scriptscriptstyle L}/T_{\scriptscriptstyle
H})^{1/2}$ \cite{NOV59,CUR73}, which is lower than the maximum
efficiency $\eta_{\mbox{\tiny C}}$ indeed. Also with two finite
thermal reservoirs as a generalization, some
Curzon-Ahlborn-Novikov-like efficiencies have independently been
achieved for the efficiency $\eta_m$, again expressed in terms of
the initially prepared temperatures of the two surroundings alone
(see, e.g., \cite{OND81,WUC88,LEE90,LEE92,OHM12,PAR14,PAR16}). All
confirm the well-known trade-off between power output and
efficiency.

Now it is worthwhile to remind, though, that in most cases of
obtaining those expressions of the efficiency $\eta_m$ so far, no
specific model of the working fluid has explicitly been in
consideration as well as, more significantly, each adiabatic process
has been simply assumed to move so fast that the temporal duration
needed for the two adiabatic processes is negligible relative to
that of the two isothermal processes. For our own purpose, let us
denote this type of methods by Approach I. Then it has been shown
with this approach that for a given pair of the finite heat source
and sink, the so-called Lorenz cycle operating in a finite time,
remarkably, exceeds the corresponding Carnot-type cycle in terms of
the efficiency $\eta_m$ \cite{LEE92,SOF93,HWA05}; the former cycle
simply replaces the two isothermal processes of the latter cycle by
the two heat-exchanging processes which are performed between two
gliding-temperature sources by adjusting the heat capacity of the
working fluid to that of the finite-capacity sources (cf. for an
extensive review for the Lorenz cycle operating either reversibly or
irreversibly, see, e.g., \cite{HWA05}).

In contrast, there has been an alternative to Approach I, denoted by
Approach II in comparison, \cite{ROS78} that the ideal-gas model of
working fluid is explicitly adopted, which enables the two finite
adiabatic times can explicitly be evaluated with the help of the
equation of state (i.e., $p V = n R T$ for the ideal gas). This more
explicit and rigorous approach, but restricted so far to the
ideal-gas model only, has applied suitably to the Carnot-type cycle
coupled to two infinite heat reservoirs, which concludes that the
(compression-ratio-independent) Curzon-Ahlborn-Novikov efficiency
can be recovered, but in the limit of an infinitely large
compression ratio of the cycle only.

In this paper, we consider the Carnot-type and the Lorenz-type
cycles operating in a finite time and then discuss the maximally
obtainable power output ${\mathcal P}_m$ and the corresponding
efficiency $\eta_m$, within a generalized form of Approach II,
explicitly being that each cycle (i) exchanges heat with the finite
heat source and sink, (ii) proceeds with a finite time for the
adiabatic processes, and (iii) explicitly employs specific gas
models for the working fluid, i.e., first the ideal-gas model then
followed by its generalization into an arbitrary real-gas model. In
doing so, we will only require the condition of endo-reversibility
that the internal relaxation time of the working fluid be so short
that it basically remains in the quasi-static state and thus
effectively satisfies the equation of state at every single instant
of time over the entire cycle (as imposed in \cite{ROS78} upon the
Carnot-type cycle operating with the ideal-gas working fluid).

We will then confirm, also within this generalized approach, the
superiority of the Lorenz cycle to the Carnot cycle in the
efficiency $\eta_m$. However, our generalized results of this
efficiency (and the maximally obtainable power output) are expressed
more realistically in terms of not only the initially prepared
temperatures of the two surroundings but also their finite heat
capacities and mass flow rates, as well as the finite heat
conductance required for the (finite-time) heat-exchanging
processes, in contrast to the results obtained in the aforecited
references. From this, it will be immediate to observe that the
Curzon-Ahlborn-Novikov results in original forms, available in the
references, can be recovered, but in the limit of an {\em
infinitely} large compression ratio and a {\em vanishingly} small
heat-exchange rate only. Therefore, the validity of the
Curzon-Ahlborn-Novikov forms within the finite thermal reservoirs
should be restricted to the case of an {\em infinitesimally} slow
change in temperatures of the two surroundings, accordingly leading
to an infinity in size of the surroundings (to cover their entire
temperature changes) and thus going back to the non-realistic setup.
As a result, it may be claimed that the endo-reversible model of
heat engine interacting with the heating and cooling fluids with a
finite heat capacity, and thus required to be confined within a
finite spatial size, should be limited in its own validity regime.

The general layout of this paper is as follows. In Sect.
\ref{sec:type-1_reservoirs} we will introduce a type of the heat
source and sink to be employed for a performance study of the
Carnot-type cycle operating in a finite time. In Sect.
\ref{sec:carnot} the maximized power output ${\mathcal P}_m$ and the
efficiency $\eta_m$ will be derived in closed form for this cycle.
In Sect. \ref{sec:type-2_reservoirs} we will introduce another type
of the heat source and sink to be used for a performance study of
the Lorenz cycle running in a finite time. In Sect. \ref{sec:lorenz}
the same analysis will take place for the latter cycle. Finally we
will give the concluding remarks of this paper in Sect.
\ref{sec:conclusion}.

%%%%%%%%%%%%%%%%%%%%%%%%%%%%%%%%%%%%%%%%%%%%%%%%%%%%%%%%%%%%%%%%%%%%%%%%
\section{Heat source and sink: Type I}\label{sec:type-1_reservoirs}
\begin{figure}[htb]
    \begin{center}
        \centering\hspace*{0cm}\vspace*{1cm}{
        \includegraphics[width=0.5\textwidth]{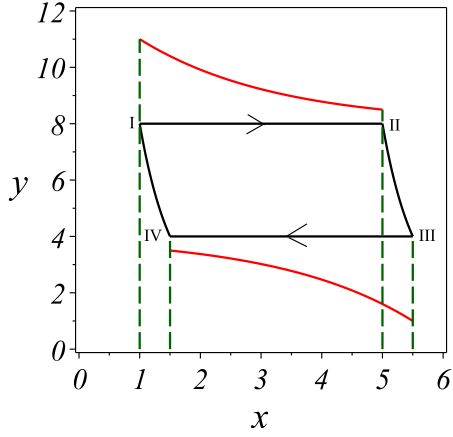}
        \caption{\label{fig:fig1}(Color online) A Carnot cycle (I $\to$ II $\to$ III $\to$ IV $\to$
        I), denoted by the closed curve (black),
        with two finite thermal reservoirs, denoted by the two outer curves (red). The $x$-axis corresponds to the volume $V$ of a working fluid in a cylinder of engine,
        and the $y$-axis to the temperature $T$. The
        evaporator is specified by $x_1 \leq x_e \leq x_2$ with
        $x_1 = 1$ and $x_2 = 5$, while the condenser by $x_4 \leq x_c \leq x_3$ with
        $x_3 = 5.5$ and $x_4 = 1.5$. The isothermal
        temperatures $(T_h^{\ast}, T_c^{\ast})$ of the working fluid
        are given here by $y = 8$ and $y = 4$, respectively. The temperatures of the two reservoirs change (``red'') as the cycle proceeds and exchanges
        heat, and so
        $T_{hi} = T_h(x_1) = 11$ and $T_{ho} = T_h(x_2) = 8.5$ as well as $T_{ci} = T_c(x_3) = 1$ and $T_{co} = T_c(x_4) = 3.5$.}}
    \end{center}
\end{figure}

%%%%%%%%%%%%%%%%%%%%%%%%%%%%%%%%%%%%%%%%%%%%%%%%%%%%%%%%%%%%%%%%%%%%%%%%
In this type, the heat source (or hot reservoir) is provided by a
heating fluid with a finite heat capacity and accordingly with its
different inlet and outlet temperatures ($T_{hi}, T_{ho})$ in the
evaporator, where $T_{hi} = T_h(x_1) > T_h(x_2) = T_{ho}$ (Fig.
\ref{fig:fig1}).
And the temperature of working fluid in the evaporator, denoted by
$T_h^{\ast}$, remains unchanged, where $T_h^{\ast} < T_{ho} <
T_{hi}$. The average rate of heat flow in the steady state from the
heat source into the engine is then given by \cite{KAY85,LEE90}
\begin{equation}\label{eq:eq1}
    \utilde{\dot{Q}}_h = \alpha_h\,(\mbox{LMTD})_h > 0\,,
\end{equation}
where the coefficient $\alpha_h$ is a (constant) heat conductance,
and the symbol $(\mbox{LMTD})_h$ denotes the logarithmic average of
temperature difference between heating fluid and working fluid such
that
\begin{equation}\label{eq:eq2}
    (\mbox{LMTD})_h = \frac{(T_{hi} - T_h^{\ast}) - (T_{ho} - T_h^{\ast})}{\ln\{(T_{hi} - T_h^{\ast})/(T_{ho} -
    T_h^{\ast})\}}\,.
\end{equation}
For simplicity, the heat capacity of heating fluid is assumed to be
a constant (i.e., an ideal-gas-like behavior). Eq. (\ref{eq:eq1})
can then be expressed as $\utilde{\dot{Q}}_h =
\dot{m}_h\,C_{ph}\,(T_{hi} - T_{ho})$ also, in terms of the mass
flow rate ($\dot{m}_h$) of the heating fluid and its heat capacity
($C_{ph}$) at constant pressure per unit mass. By combining these
two expressions for $\utilde{\dot{Q}}_h$, we can easily determine
the outlet temperature of evaporator as
\begin{equation}\label{eq:temperature-h2_1}
    T_{ho} = T_h^{\ast} + (T_{hi} - T_h^{\ast})\,
    e^{-A_h}\,,
\end{equation}
expressed in terms of the initially prepared temperatures $T_{hi}$
and $T_h^{\ast}$, as well as the dimensionless quantity $A_h =
\alpha_h/(\dot{m}_h\,C_{ph})$.

Similarly, the heat sink (or cold reservoir) is given by a cooling
fluid with its inlet and outlet temperatures ($T_{ci}, T_{co}$) in
the condenser, where $T_{ci} = T_c(x_3) < T_c(x_4) = T_{co}$. Then
the average rate of heat flow from the engine to the heat sink is
\begin{equation}\label{eq:eq3}
    \utilde{\dot{Q}}_c = \alpha_c\,(\mbox{LMTD})_c \stackrel{!}{=} \dot{m}_c\,C_{pc}\,(T_{co} - T_{ci}) > 0\,,
\end{equation}
where the logarithmic average of temperature difference between
cooling fluid and working fluid,
\begin{equation}\label{eq:eq4}
    (\mbox{LMTD})_c = \frac{(T_c^{\ast} - T_{ci}) - (T_c^{\ast} - T_{co})}{\ln\{(T_c^{\ast} - T_{ci})/(T_c^{\ast} -
    T_{co})\}}
\end{equation}
with the condenser temperature $T_c^{\ast}$ of working fluid, where
$T_c^{\ast} > T_{co}
> T_{ci}$. Eq. (\ref{eq:eq3}) can easily determine the outlet
temperature of condenser as
\begin{equation}\label{eq:temperature-c2_1}
    T_{co} = T_c^{\ast} + (T_{ci} - T_c^{\ast})\, e^{-A_c}\,,
\end{equation}
where $A_c = \alpha_c/(\dot{m}_c\,C_{pc})$.

Now we are to complete the temperature profile of heating fluid.
This temperature monotonically decreases with a duration of heat
exchange, and the ratio of this temperature change to the change of
position ($x_e$) in the evaporator, with $x_1 \leq x_e \leq x_2$
(Fig. \ref{fig:fig1}), should be proportional to the temperature
difference between heating fluid and working fluid at the very
position. This gives
\begin{equation}\label{eq:heat-transport_1}
    \frac{d T_h}{d x_e} = -k_h\,\{T_h(x_e) - T_h^{\ast}\}\,,
\end{equation}
where the proportionality constant $k_h$ will uniquely be determined
below; if $k_h =0$, this heat source simply reduces to an infinite
thermal reservoir. Integrating Eq. (\ref{eq:heat-transport_1}) over
$x_e$ will easily yield the exponentially decaying temperature
\begin{equation}\label{eq:temperature-heat-source_1}
    T_h(x_e) = T_h^{\ast} + (T_{hi} - T_h^{\ast})\,e^{-k_h (x_e-x_1)}\,.
\end{equation}
Setting $x_e = x_2$ and then equating this to Eq.
(\ref{eq:temperature-h2_1}), we can finally determine the
proportionality as
\begin{equation}\label{eq:coefficient_1}
    k_h = \frac{A_h}{x_2 - x_1}\,.
\end{equation}

Likewise, we have, for the temperature profile of cooling fluid,
\begin{equation}\label{eq:heat-transport_2}
    \frac{d T_c}{d x_c} = -k_c\,\{T_c^{\ast} - T_c(x_c)\}\,,
\end{equation}
where the position in the condenser, $x_4 \leq x_c \leq x_3$. This
easily gives
\begin{equation}\label{eq:temperature-heat-sink_1-1}
    T_c(x_c) = T_c^{\ast} - (T_c^{\ast} - T_{ci})\,e^{-k_c
    (x_3-x_c)}\,,
\end{equation}
which can, together with Eq. (\ref{eq:temperature-c2_1}), yield
\begin{equation}\label{eq:coefficient_1-1}
    k_c = \frac{A_c}{x_3 - x_4}\,.
\end{equation}

%%%%%%%%%%%%%%%%%%%%%%%%%%%%%%%%%%%%%%%%%%%%%%%%%%%%%%%%%%%%%
\section{Carnot-type cycle operating in a finite time}\label{sec:carnot}
%%%%%%%%%%%%%%%%%%%%%%%%%%%%%%%%%%%%%%%%%%%%%%%%%%%%%%%%%%%%%
The system in consideration consists of a working fluid in a
cylinder attached to a piston. The working fluid will be given by an
arbitrary model of the real gas, which satisfies the expression of
internal energy \cite{CAS83,TJI06}
\begin{equation}\label{eq:1st-law_1}
    dU = C_{\scriptscriptstyle V}\,dT + \left\{T\,\left(\frac{\partial p}{\partial T}\right)_{\scriptscriptstyle V} -
    p\right\}\,dV\,.
\end{equation}
From comparison with the First Law $dU = \delta Q - p\,dV$, the
infinitesimal heat is then
\begin{equation}\label{eq:1st-law_2}
    \delta Q = C_{\scriptscriptstyle V}\,dT + T\,\left(\frac{\partial p}{\partial T}\right)_{\scriptscriptstyle
    V}\,dV\,,
\end{equation}
which will be employed below. For simplicity, we will begin with the
ideal-gas model for the working fluid.

%%%%%%%%%%%%%%%%%%%%%%%%%%%%%%%%%%%%%%%%%%%%%%%%%%%%%%%%
\subsection{Isothermal Expansion}\label{sec:isothermal}
%%%%%%%%%%%%%%%%%%%%%%%%%%%%%%%%%%%%%%%%%%%%%%%%%%%%%%%%
In this process of heat exchange between heating fluid and working
fluid, the rate of heat exchange satisfies
\begin{equation}\label{eq:eq5}
    \dot{Q}_h(V) = \alpha_h\,\{T_h(V) - T_h^{\ast}\} \stackrel{!}{=}
    p\,\frac{dV}{dt} > 0
\end{equation}
[cf. (\ref{eq:1st-law_2})], in which the volume of working fluid
satisfies $V(x_1) \leq V(x_e) \leq V(x_2)$. Let $V = {\mathcal
A}\,x_e$ where the symbol ${\mathcal A}$ denotes the (constant)
cross sectional area of the cylinder. Substituting $p = n
R\,T_{h}^{\ast}/V$ into Eq. (\ref{eq:eq5}) easily gives rise to
\begin{equation}\label{eq:eq6}
    \frac{dt}{\tau_h} = \frac{dV}{V} = \frac{dx_e}{x_e}\,,
\end{equation}
where the symbol $\tau_h(x_e) = n
R\,T_h^{\ast}\cdot(\alpha_h)^{-1}\,\{T_h(x_e) - T_h^{\ast}\}^{-1}$.
Subsequently, we substitute Eq. (\ref{eq:temperature-heat-source_1})
into this symbol and then integrate (\ref{eq:eq6}) over $x_e$, the
duration of this isothermal process can be determined as
\begin{equation}\label{eq:eq7}
    (t_h)_{\mbox{\scriptsize id}} = \tau_h(x_1)\,e^{-A_{h1}}\,\left\{\mbox{Ei}(A_{h1}\,r_{12}) -
    \mbox{Ei}(A_{h1})\right\}\,,
\end{equation}
where the dimensionless quantity $A_{h1} = A_h\,x_1/(x_2-x_1)$, and
the ratio $r_{12} = x_2/x_1$, as well as the exponential integral
\cite{ABR65}
\begin{equation}\label{eq:eq8}
    \mbox{Ei}(y) = P \int_{-\infty}^y \frac{\exp(s)}{s}\,ds
\end{equation}
with $P$ denoting Cauchy's principal value for $y > 0$. This
exponential integral can also be rewritten as a series expansion
\begin{equation}\label{eq:series_1}
    \mbox{Ei}(y) = \gamma_e + \ln(y) + \sum_{n=1}^{\infty} \frac{y^n}{n\cdot
    n!}\,,
\end{equation}
in which Euler's constant $\gamma_e = 0.5772$. In case that $C_{ph}
\to \infty$ and so $A_{h1} \to 0$, then it is straightforward to
observe, with the help of (\ref{eq:series_1}), that
\begin{equation}\label{eq:inf-thermal-reservoir-case-1}
    (t_h)_{\mbox{\scriptsize id},\infty} = \tau_h(x_1)\,\ln(r_{12})\,,
\end{equation}
which is identical to the expression obtained in \cite{ROS78}. As a
result, the total amount of heat input can explicitly be computed,
with the help of Eqs. (\ref{eq:eq5}), (\ref{eq:eq6}) and
(\ref{eq:inf-thermal-reservoir-case-1}), as
\begin{equation}\label{eq:heat-input_1}
    (Q_h)_{\mbox{\scriptsize id}} = \int_0^{t_h} \dot{Q}_h\{x_e(t')\}\,dt' = n R\,T_{h}^{\ast}\,\ln(r_{12})\,,
\end{equation}
which is also valid for $C_{ph} \to \infty$ indeed.

Next we repeat the afore-developed discussion, but within an
arbitrary model of the real gas as a generalization. With the help
of (\ref{eq:1st-law_2}), Eq. (\ref{eq:eq5}) should then be replaced
by
\begin{equation}\label{eq:heat-rate-arbitrary_1}
    \dot{Q}_h = \alpha_h\,\{T_h(x_e) - T_h^{\ast}\}
    \stackrel{!}{=} T_h^{\ast}\,\left(\frac{\partial p}{\partial T}\right)_{\scriptscriptstyle
    V}\,\frac{dV}{dt} > 0\,.
\end{equation}
This will easily yield the total amount of heat input
\begin{equation}\label{eq:heat-arbitrary_1}
    Q_h = R\,T_h^{\ast}\,\{F(V_2, T_h^{\ast}) - F(V_1,
    T_h^{\ast})\}\,,
\end{equation}
in which the dimensionless quantity $F(V, T) := R^{-1}
\int^{\scriptscriptstyle V} (\partial p/\partial
T)_{\scriptscriptstyle V'}\,dV'$. As two specific examples of the
real gas, the van der Waals gas, with the equation of state given by
$p = n R\,T/(V-nb) - n^2 a/V^2$ \cite{BER07}, yields
$(F)_{\mbox{\tiny vdW}} = n\,\ln(V-nb)$, and the Redlich-Kwong gas,
with $p = n R\,T/(V-nb) - n^2 a/\{T^{1/2}\,V\,(V+nb)\}$
\cite{WAR77}, gives $(F)_{\mbox{\tiny RK}} = n\,\ln(V-nb) - \{n a/(2
b\,R\,T^{3/2})\}\,\ln\{(V+nb)/V\}$.

The duration of isothermal expansion, too, can be determined
explicitly. From (\ref{eq:heat-rate-arbitrary_1}), it follows that
\begin{equation}\label{eq:time-duration-1}
    \frac{dt}{\tau_h(x)} = \frac{1}{n R}\, \left(\frac{\partial p}{\partial
    T}\right)_x^{\scriptscriptstyle T\to T_h^{\ast}}\,dx
\end{equation}
[cf. (\ref{eq:eq6})], which subsequently gives, similarly to
(\ref{eq:eq7}), the duration
\begin{equation}\label{eq:time-duration-2}
    t_h = \frac{\tau_h(x_1)}{n R}\,e^{-A_{h1}}\,\int_{x_1}^{x_2} e^{k_h
    x}\,\left(\frac{\partial p}{\partial T}\right)_x^{\scriptscriptstyle T\to T_h^{\ast}}\,dx\,.
\end{equation}
It is easy to confirm that in the limit of $C_{ph} \to \infty$, this
reduces to
\begin{equation}\label{eq:time-duration-2-1}
    (t_h)_{\infty} = \frac{\tau_h(x_1)}{n}\,\{F(V_2, T_h^{\ast}) - F(V_1, T_h^{\ast})\}\,.
\end{equation}
For the van der Waals gas, Eq. (\ref{eq:time-duration-2}) can
explicitly be evaluated as the closed form
\begin{equation}\label{eq:eq7-vdW}
    (t_h)_{\mbox{\tiny vdW}} = \tau_h(x_1)\,e^{-A_{h1}}\,\Lambda_h(\tilde{b})\,,
\end{equation}
where the symbol $\Lambda_h(\tilde{b}) := e^{k_h
\tilde{b}}\,\{\mbox{Ei}(\tilde{A}_{h1}\,\tilde{r}_{12}) -
\mbox{Ei}(\tilde{A}_{h1})\}$ with $\tilde{b} = n b/{\mathcal A}$ and
$\tilde{r}_{12} = (x_2-\tilde{b})/(x_1-\tilde{b})$, as well as
$\tilde{A}_{h1} = A_h\,(x_1 - \tilde{b})/(x_2-x_1)$. For the
Redlich-Kwong gas, we have
\begin{equation}\label{eq:eq7-RK}
    (t_h)_{\mbox{\tiny RK}} = (t_h)_{\mbox{\tiny vdW}} + \tau_h(x_1)\,e^{-A_{h1}}\,\frac{a}{2 b\,R\,(T_h^{\ast})^{3/2}}\,
    \left\{\Lambda_h(0) - \Lambda_h(-\tilde{b})\right\}\,.
\end{equation}

For a later purpose, it is also interesting to consider the
asymptotic behavior of the isothermal duration in
(\ref{eq:time-duration-2}), in the limit of $r_{12} = x_2/x_1 \to
\infty$, with the help of the expansion
\begin{equation}\label{eq:time-duration-3}
    e^{k_h x} = 1 + \sum_{n=1}^{\infty} \left(\frac{\alpha_h}{\dot{m}_h\,C_{ph}}\,\frac{x}{x_2 -
    x_1}\right)^n\,.
\end{equation}
Due to the fact that the leading term of pressure $p$ is, for a
generic real gas, given by $\propto (V \pm b)^{-1}\,\chi(T) \propto
(x \pm \tilde{b})^{-1}\,\chi(T)$ with a certain generic function
$\chi(T)$ \cite{CAS83}, it is straightforward to observe that in the
limit of $r_{12} \to \infty$, all terms with $n \geq 1$ on the
right-hand side of (\ref{eq:time-duration-3}) will not make a
non-vanishing contribution to the integrand on the right-hand side
of (\ref{eq:time-duration-2}). As a result, the duration of
isothermal expansion exactly reduces, with $r_{12} \to \infty$, to
the same expression as (\ref{eq:time-duration-2-1}). Combining this
expression with (\ref{eq:heat-arbitrary_1}), the average rate of
heat input is easily given by the universal form
\begin{equation}\label{eq:time-duration-4-1}
    \utilde{\dot{Q}}_h = \frac{Q_h}{t_h} \to (\utilde{\dot{Q}}_h)_{\infty} = \frac{n R\,T_h^{\ast}}{\tau_h(x_1)} = \alpha_h\,\{T_h(x_1) -
    T_h^{\ast}\}\,,
\end{equation}
valid for an arbitrary gas model of the working fluid, but in the
limit of either $C_{ph} \to \infty$ or $r_{12} \to \infty$ only.

%%%%%%%%%%%%%%%%%%%%%%%%%%%%%%%%%%%%%%%%%%%%%%%%%%%%%%%
\subsection{Adiabatic Expansion}\label{sec:adiabatic}
%%%%%%%%%%%%%%%%%%%%%%%%%%%%%%%%%%%%%%%%%%%%%%%%%%%%%%%
Because there is no heat exchange ($\delta Q = 0$) in this process
[cf. (\ref{eq:1st-law_2})], the duration of adiabatic expansion with
a finite thermal reservoir should be the same as that obtained
within an infinite reservoir. Due to the non-availability of a
constraint, corresponding to (\ref{eq:eq5}) valid for an isothermal
process, which will lead to an explicit evaluation of the adiabatic
duration, it is simply reasonably assumed \cite{CUR73,ROS78} that
this duration is proportional to that of isothermal expansion with
$dx/dt = x/\tau_h$ [cf. (\ref{eq:eq6}) and
(\ref{eq:inf-thermal-reservoir-case-1})]. Consequently, we acquire
the duration for the ideal gas
\begin{equation}\label{eq:adiabatic-expansion_1}
    (t_{a1})_{\mbox{\scriptsize id}} = \tau_h(x_1)\, \ln\left(\frac{x_3}{x_2}\right) \stackrel{!}{=} \frac{\tau_h(x_1)}{\gamma -
    1}\, \ln\left(\frac{T_h^{\ast}}{T_c^{\ast}}\right)\,.
\end{equation}
Here for the second equality we applied
$T_h^{\ast}\,(V_2)^{\gamma-1} = T_c^{\ast}\,(V_3)^{\gamma-1}$, valid
during an adiabatic process, where $\gamma =
(C_p)_{\mbox{\scriptsize id}}/(C_{\scriptscriptstyle
V})_{\mbox{\scriptsize id}}$ is the ratio of the heat capacity at
constant pressure to heat capacity at constant volume, with
$(C_p)_{\mbox{\scriptsize id}}\,-\,(C_{\scriptscriptstyle
V})_{\mbox{\scriptsize id}} = n R$.

Subsequently, an arbitrary real gas is under consideration. We then
employ the infinite-reservoir result given in
(\ref{eq:time-duration-2-1}) and obtain the duration of adiabatic
expansion
\begin{equation}\label{eq:time-duration-2-1-adiabatic}
    t_{a1} = \frac{\tau_h(x_1)}{n}\,\{F(V_3, T_c^{\ast}) - F(V_2, T_h^{\ast})\} \stackrel{!}{=} \frac{\tau_h(x_1)}{n R}\,\{g(T_h^{\ast}) - g(T_c^{\ast})\} >
    0\,,
\end{equation}
where $g(T) = \int^{\scriptscriptstyle T} \{f(T')/T'\}\,dT'$ is a
function of temperature only, with a function $f(T)$ satisfying the
expression of heat capacity \cite{TJI06}
\begin{equation}\label{eq:C_v-arbitrary-gas_1-0}
    C_{\scriptscriptstyle V} = T\,\int^{\scriptscriptstyle V} \left(\frac{\partial^2 p}{\partial
    T^2}\right)_{\scriptscriptstyle V'}\,dV' + f(T)\,,
\end{equation}
and $C_p - C_{\scriptscriptstyle V} = -T\,\{(\partial p/\partial
T)_{\scriptscriptstyle V}\}^2/(\partial p/\partial
V)_{\scriptscriptstyle T} > 0$ \cite{BER07}. The second equality of
Eq. (\ref{eq:time-duration-2-1-adiabatic}) is proved in Appendix. In
the ideal gas, $f(T) \to (C_{\scriptscriptstyle
V})_{\mbox{\scriptsize id}} = (d/2)\,n R$ with the number of degrees
of freedom $d$, which enables (\ref{eq:time-duration-2-1-adiabatic})
to reduce to (\ref{eq:adiabatic-expansion_1}) indeed. In the van der
Waals gas and the Redlich-Kwong gas (also even in an arbitrary gas
model), we may set $f(T) \to (C_{\scriptscriptstyle
V})_{\mbox{\scriptsize id}}$, which enables
(\ref{eq:time-duration-2-1-adiabatic}) to reduce to
(\ref{eq:adiabatic-expansion_1}), too.

%%%%%%%%%%%%%%%%%%%%%%%%%%%%%%%%%%%%%%%%%%%%%%%%%%%%%%%%%%%%%
\subsection{Isothermal Compression}\label{sec:iso-thermal}
%%%%%%%%%%%%%%%%%%%%%%%%%%%%%%%%%%%%%%%%%%%%%%%%%%%%%%%%%%%%%
The rate of heat exchange in the condenser is, for the ideal gas,
given by
\begin{equation}\label{eq:eq5-1}
    \dot{Q}_c(V) = \alpha_c\,\{T_c(V) - T_c^{\ast}\} \stackrel{!}{=}
    p\,\frac{dV}{dt} < 0\,,
\end{equation}
in which the volume of working fluid satisfies $V(x_3) \geq V(x_c)
\geq V(x_4)$. Substituting $p = n R\,T_{c}^{\ast}/V$ into
(\ref{eq:eq5-1}) gives rise to
\begin{equation}\label{eq:eq6-1}
    \frac{dt}{\tau_c} = -\frac{dV}{V} = -\frac{dx_c}{x_c}\,,
\end{equation}
where $\tau_c(x_c) = n
R\,T_c^{\ast}\cdot(\alpha_c)^{-1}\,\{T_c^{\ast} - T_c(x_c)\}^{-1} >
0$. Plugging (\ref{eq:temperature-heat-sink_1-1}) into this and then
integrating (\ref{eq:eq6-1}) over $x_c$ will yield
\begin{equation}\label{eq:eq7-1}
    (t_c)_{\mbox{\scriptsize id}} = \tau_c(x_3)\,e^{A_{c3}}\,\left\{\mbox{E}_1(A_{c3}/r_{34}) -
    \mbox{E}_1(A_{c3})\right\}\,,
\end{equation}
where the dimensionless quantity $A_{c3} = A_c\,x_3/(x_3-x_4)$, and
the ratio $r_{34} = x_3/x_4$, as well as the exponential integral
\cite{ABR65}
\begin{equation}\label{eq:exponential_integral_2}
    \mbox{E}_1(y) = \int_y^{\infty} \frac{\exp(-s)}{s}\,ds
\end{equation}
where $\mbox{E}_1(y) = \Gamma(0,y)$ with the incomplete gamma
function $\Gamma(a,y)$ and $y > 0$, also expressed as a series
expansion
\begin{equation}\label{eq:series_2}
    \mbox{E}_1(y) = -\gamma_e - \ln(y) - \sum_{n=1}^{\infty}
    \frac{(-1)^n\,y^n}{n\,n!}\,.
\end{equation}
With the help of (\ref{eq:eq5-1}) and (\ref{eq:eq7-1}), the total
amount of heat exchange becomes
\begin{equation}\label{eq:heat-input_1-1}
    |(Q_c)_{\mbox{\scriptsize id}}| = -\int_0^{t_c} \dot{Q}_c\{x_c(t')\}\,dt' = n
    R\,T_{c}^{\ast}\,\ln(r_{12})\,,
\end{equation}
where we used the relation $x_3/x_4 = x_2/x_1$. For a later purpose,
it is also instructive to note that if $C_{pc} \to \infty$ and so
$A_{c3} \to 0$, then Eq. (\ref{eq:eq7-1}) reduces, with the help of
(\ref{eq:series_2}), to its limiting value
\begin{equation}\label{eq:inf-thermal-reservoir-case-1-0}
    (t_c)_{\mbox{\scriptsize id},\infty1} =
    \tau_c(x_3)\,\ln(r_{12})\,,
\end{equation}
while if $r_{34} = r_{12} \to \infty$, then we get $e^{A_{c3}} \to
e^{A_c}$ and so Eq. (\ref{eq:eq7-1}) reduces to a different limiting
value
\begin{equation}\label{eq:inf-thermal-reservoir-case-1-1}
    (t_c)_{\mbox{\scriptsize id},\infty2} = e^{A_c}\,(t_c)_{\mbox{\scriptsize id},\infty1}\,.
\end{equation}

Next we consider an arbitrary real-gas model for the working fluid.
Based on the justification similar to that used for the isothermal
expansion, it is straightforward to obtain the total amount of heat
exchange in the isothermal compression
\begin{equation}\label{eq:heat-arbitrary_1-0}
    |Q_c| = R\,T_c^{\ast}\,\{F(V_3, T_c^{\ast}) - F(V_4, T_c^{\ast})\} = R\,T_c^{\ast}\,\{F(V_2, T_h^{\ast}) - F(V_1, T_h^{\ast})\}\,,
\end{equation}
where a proof of the second equality is provided in Appendix. The
duration of isothermal compression will then be
\begin{equation}\label{eq:time-duration-cold_1}
    t_c = \frac{\tau_c(x_3)}{n R}\,e^{A_{c3}}\,\int_{x_4}^{x_3}
    e^{-k_c x}\,\left(\frac{\partial p}{\partial T}\right)_x^{\scriptscriptstyle T\to T_c^{\ast}}\,dx\,.
\end{equation}
In the limit of $C_{ph} \to \infty$, this reduces to
\begin{equation}\label{eq:time-duration-2-1-0}
    (t_c)_{\infty1} = \frac{\tau_c(x_3)}{n}\,\{F(V_2, T_h^{\ast}) - F(V_1,
    T_h^{\ast})\}
\end{equation}
[cf. (\ref{eq:inf-thermal-reservoir-case-1-0})], leading the average
rate of heat exchange, with the aid of
(\ref{eq:heat-arbitrary_1-0}), to
\begin{equation}\label{eq:time-duration-4-1-0}
    \utilde{\dot{Q}}_c = \frac{|Q_c|}{t_c} \to (\utilde{\dot{Q}}_c)_{\infty1} = \frac{n R\,T_c^{\ast}}{\tau_c(x_3)} = \alpha_c\,\{T_c^{\ast} - T_c(x_3)\}\,.
\end{equation}
In the limit of $r_{34} \to \infty$, in contrast, it follows that
[cf. (\ref{eq:inf-thermal-reservoir-case-1-1})]
\begin{eqnarray}
    (t_c)_{\infty2} &=& e^{A_c}\,(t_c)_{\infty1}\label{eq:time-duration-cold-2}\\
    (\utilde{\dot{Q}}_c)_{\infty2} &=&
    e^{-A_c}\,(\utilde{\dot{Q}}_c)_{\infty1}\,.\label{eq:time-duration-4-1-1}
\end{eqnarray}
From (\ref{eq:time-duration-cold_1}) we can obtain the closed form
for the van der Waals gas
\begin{equation}\label{eq:eq7-vdW-0}
    (t_c)_{\mbox{\tiny vdW}} = \tau_c(x_3)\,e^{A_{c3}}\,\Lambda_c(\tilde{b})\,,
\end{equation}
where $\Lambda_c(\tilde{b}) := e^{-k_c
\tilde{b}}\,\{\mbox{E}_1(\tilde{A}_{c3}/\tilde{r}_{34}) -
\mbox{E}_1(\tilde{A}_{c3})\}$ with $\tilde{r}_{12} = \tilde{r}_{34}
= (x_3-\tilde{b})/(x_4-\tilde{b})$ and $\tilde{A}_{c3} = A_c\,(x_3 -
\tilde{b})/(x_3-x_4)$. For the Redlich-Kwong gas,
\begin{equation}\label{eq:eq7-RK-0}
    (t_c)_{\mbox{\tiny RK}} = (t_c)_{\mbox{\tiny vdW}} + \tau_c(x_3)\,e^{A_{c3}}\,\frac{a}{2 b\,R\,(T_c^{\ast})^{3/2}}\,
    \left\{\Lambda_c(0) - \Lambda_c(-\tilde{b})\right\}\,.
\end{equation}

%%%%%%%%%%%%%%%%%%%%%%%%%%%%%%%%%%%%%%%%%%%%%%%%%%%%%%%%%%%%%%
\subsection{Adiabatic Compression}\label{sec:adiabatic-2}
%%%%%%%%%%%%%%%%%%%%%%%%%%%%%%%%%%%%%%%%%%%%%%%%%%%%%%%%%%%%%%
Following to the same method as applied to the adiabatic expansion,
the duration of adiabatic compression is given by
\begin{equation}\label{eq:time-duration-2-1-adiabatic-20}
    t_{a2} = \frac{\tau_c(x_3)}{n}\,\{F(V_4, T_c^{\ast}) - F(V_1, T_h^{\ast})\}
    \stackrel{!}{=}
    \frac{\tau_c(x_3)}{n R}\,\{g(T_h^{\ast}) - g(T_c^{\ast})\} > 0\,,
\end{equation}
which reduces, in the ideal-gas model, to
\begin{equation}\label{eq:adiabatic-expansion_1-20i}
    (t_{a2})_{\mbox{\scriptsize id}} = \tau_c(x_3)\, \ln\left(\frac{x_3}{x_2}\right) \stackrel{!}{=} \frac{\tau_c(x_3)}{\gamma - 1}\, \ln\left(\frac{T_h^{\ast}}{T_c^{\ast}}\right)\,.
\end{equation}
Along the similar lines to the adiabatic expansion, we may finally
set that Eq. (\ref{eq:adiabatic-expansion_1-20i}) is also valid for
the van der Waals gas and the Redlich-Kwong gas.

As a result, the duration of a single cycle is
\begin{equation}\label{eq:total-time-of-a-single-cycle_1}
    t_{cy} = t_h + t_{a1} + t_c + t_{a2} = (t_h + t_c) (1 +
    \xi)
\end{equation}
[cf. (\ref{eq:time-duration-2}),
(\ref{eq:time-duration-2-1-adiabatic}),
(\ref{eq:time-duration-cold_1}) and
(\ref{eq:time-duration-2-1-adiabatic-20})], where $\xi = (t_{a1} +
t_{a2})/(t_h + t_c)$. To neglect the total adiabatic duration, the
condition of $\xi \ll 1$ must be met.

%%%%%%%%%%%%%%%%%%%%%%%%%%%%%%%%%%%%%%%%%%%%%%%%%%%%%%
\subsection{Power output and its optimization}\label{sec:power-output}
%%%%%%%%%%%%%%%%%%%%%%%%%%%%%%%%%%%%%%%%%%%%%%%%%%%%%%
The power output of Carnot cycle is given, in the ideal-gas model,
by
\begin{equation}\label{eq:power-output_1}
    {\mathcal P} = \frac{Q_h - Q_c}{t_{cy}} = \frac{n R\,(T_h^{\ast} -
    T_c^{\ast})\,\ln(r_{12})}{t_{cy}}\,,
\end{equation}
where $W = Q_h - Q_c$. For the sake of simplicity, let $\alpha_h =
\alpha_c =: \alpha$ and $C_{ph} = C_{pc} =: C_p$ as well as
$\dot{m}_h = \dot{m}_c =: \dot{m}$, and so $A_h = A_c =: A$ and
$A_{h1} =: A_1$. We introduce the dimensionless quantities such as
$w = T_h^{\ast}/T_{hi}$ and $z = T_c^{\ast}/T_h^{\ast}$ as well as
$\beta = T_{ci}/T_{hi}$, which can rewrite Eq.
(\ref{eq:power-output_1}) as
\begin{equation}\label{eq:power-output_2}
    {\mathcal P} = \frac{\alpha\,T_{hi}\,(1 - z)\,(1  + \lambda\,\ln
    z)\,\ln r}{\mbox{Den}}\,,
\end{equation}
in which the compression ratio $r = x_3/x_1 = r_{12}\,z^{-\nu}$ with
$\nu = (\gamma-1)^{-1}$, and $\lambda := \nu/(\ln r)$, as well as
the denominator
\begin{eqnarray}\label{eq:denominator_1}
    \mbox{Den} &=& \frac{1}{1-w}\,\left[e^{-A_1}\,\left\{\mbox{Ei}(A_1\,r_{12}) -
    \mbox{Ei}(A_1)\right\} - \nu\,\ln z\right] +\n\\
    && \frac{z}{w z - \beta}\,\left[e^{A_1\,r_{12}}\,\left\{\mbox{E}_1(A_1) -
    \mbox{E}_1(A_1\,r_{12})\right\} - \nu\,\ln z\right]\,.
\end{eqnarray}

Now the maximization of ${\mathcal P}(w, z)$ with respect to $w$ and
$z$ (i.e., $T_h^{\ast}$ and $T_c^{\ast}$) is under consideration.
First, by requiring $\partial {\mathcal P}/\partial w
\stackrel{!}{=} 0$ at $w = w_m$ and $z = z_m$ for the maximum value
${\mathcal P}_m$, we can obtain
\begin{eqnarray}\label{eq:optimization_1}
    && (w_m\,z_m - \beta)^2\,\left[e^{-A_1}\,\left\{\mbox{Ei}(A_1\,r_{12}) -
    \mbox{Ei}(A_1)\right\} - \nu\,\ln z_m\right]\n\\
    &\stackrel{!}{=}& z_m^2\,(1-w_m)^2\,\left[e^{A_1\,r_{12}}\,\left\{\mbox{E}_1(A_1) -
    \mbox{E}_1(A_1\,r_{12})\right\} - \nu\,\ln z_m\right]\,.
\end{eqnarray}
Next, from the requirement of $\partial {\mathcal P}/\partial z
\stackrel{!}{=} 0$, we can also find another equality similar to
(\ref{eq:optimization_1}), denoted by (\ref{eq:optimization_1})$'$,
but too complicated in form and thus not explicitly given here. It
is now instructive to consider a simple case of  $A_1 =
\alpha/(\dot{m}\,C_p) \to 0$. With the help of (\ref{eq:series_1})
and (\ref{eq:series_2}), Eq. (\ref{eq:optimization_1}) then reduces
to
\begin{equation}\label{eq:optimization_2}
    (z_m-\beta)\,\{(2 w_m - 1)\,z_m - \beta\}\,\ln r = 0\,.
\end{equation}
Because $z_m = \beta$ simply means a quasi-static process, we should
instead choose $w_m = (z_m + \beta)/(2 z_m)$, as obtained in
\cite{ROS78} in the case of $C_p \to \infty$. Substituting this
expression of $w_m$ into (\ref{eq:optimization_1})$'$ with $C_p \to
\infty$ (and so $A_1 \to 0$) has actually led to the fact that the
Curzon-Ahlborn-Novikov result $\eta_{\mbox{\tiny CAN}} = 1 - z_m$
with $z_m = \beta^{1/2} + {\mathcal O}(\lambda^1)$ can be recovered
indeed, but in the limit of $r \to \infty$ only (even for the ideal
gas!) [cf. Eqs. (A5)-(A8) of \cite{ROS78} an explicit expression of
the compression-ratio dependent higher-order terms ${\mathcal
O}(\lambda^1)$]. Within the finite thermal reservoir $C_p
\nrightarrow \infty$, in contrast, the case of $A_1 \to 0$
corresponds to a vanishingly small heat conductance $\alpha \to 0$,
or a quasi-static heat exchange, which necessarily requires that
$z_m = \beta$ in (\ref{eq:optimization_2}). This suggests that
within the finite thermal reservoirs, it is not straightforward to
observe the Curzon-Ahlborn-Novikov result, even for the ideal gas in
the limit of an infinitely large compression ratio. Figs. 2 and 3
demonstrate various behaviors of ${\mathcal P}_m$ and $\eta_m$
versus the ratio $r_{12}$ (understood as the lower bound of the
compression ratio $r$), respectively, for the ideal gas, the van der
Waals gas, and the Redlich-Kwong gas. We then observe that the
maximum power output ${\mathcal P}_m$ increases with $r_{12}$ (or
$r$), and its maximum, denoted by $({\mathcal P}_m)_m$, will be
achieved in the limit of $r \to \infty$.
\begin{figure}[htb]
    \begin{center}
        \centering\hspace*{0cm}\vspace*{1cm}{
        \includegraphics[width=0.5\textwidth]{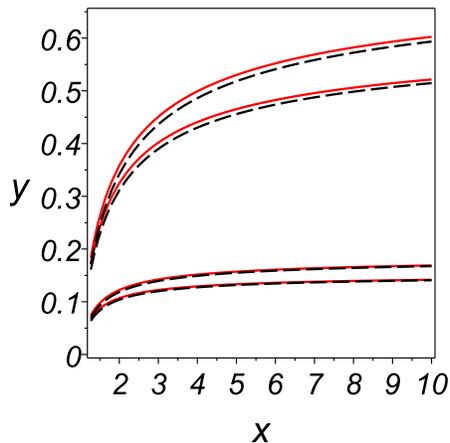}
        \caption{\label{fig:fig2}(Color online) The maximally obtainable power output of the Carnot
        cycle normalized by the Curzon-Ahlborn-Novikov power output,
        $y = {\mathcal P}_m/{\mathcal P}_{\mbox{\tiny CAN}}$ versus
        $x = r_{12}$. The power output ${\mathcal P} = (Q_h -
        Q_c)/t_{cy}$ with (\ref{eq:total-time-of-a-single-cycle_1}), and ${\mathcal P}_{\mbox{\tiny CAN}}$
        in (\ref{eq:power-output-max-reservoir_1}), here with $\beta = 0.27$.
        We used Eqs. (\ref{eq:heat-arbitrary_1}), (\ref{eq:time-duration-2}), (\ref{eq:adiabatic-expansion_1}),
        (\ref{eq:heat-arbitrary_1-0}),
        (\ref{eq:time-duration-cold_1}) and (\ref{eq:adiabatic-expansion_1-20i}) with $\gamma = 1.4$, as well as the equations of motion [cf. after
        (\ref{eq:heat-arbitrary_1})], with (\ref{eq:power-output_2}) and
        (\ref{eq:denominator_1}) for the ideal gas, and with the aid of (\ref{eq:eq7-vdW}) and
        (\ref{eq:eq7-vdW-0}) for the van der Waals gas, as well as with the aid of (\ref{eq:eq7-RK}) and (\ref{eq:eq7-RK-0}) for the Redlich-Kwong
        gas.
        The maximization of ${\mathcal P}$ was carried out by means of {\em Maple} to evaluate $(w_m, z_m)$.
        The solid and red curves are given for the ideal gas, and the dash
        and black ones for the van der Waals gas with $\tilde{b}/x_1 = 0.1$. From top to bottom: $(\beta, A) = (0.27, 0.1); (0.27, 0.5); (0.6,
        0.1); (0.6, 0.5)$. The curves for the Redlich-Kwong gas with $\tilde{b}/x_1 = 0.1$ and $a/(2 b\,R\,T^{3/2}) \leq 0.1$ are not distinguished
        from their van-der-Waals counterparts.}}
    \end{center}
\end{figure}
\begin{figure}[htb]
    \begin{center}
        \centering\hspace*{0cm}\vspace*{1cm}{
        \includegraphics[width=0.5\textwidth]{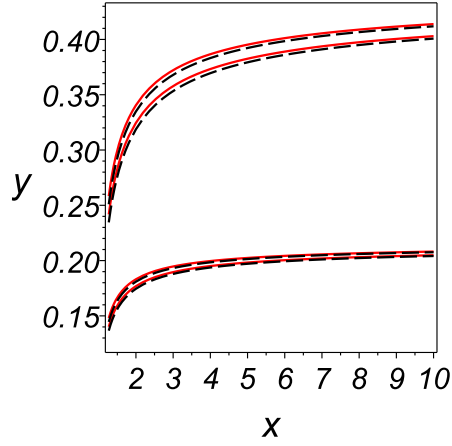}
        \caption{\label{fig:fig3}(Color online) The efficiency, $y = \eta_m$, at the point of ${\mathcal
        P}_m$ plotted in Fig. \ref{fig:fig2} versus $x = r_{12}$.
        Here $\eta_m = 1 - z_m$.
        The solid and red curves are given for the ideal gas, and the dash
        and black ones for the van der Waals gas with $\tilde{b}/x_1 = 0.1$. From top to bottom: $(\beta, A) = (0.27, 0.5); (0.27, 0.1); (0.6,
        0.5); (0.6, 0.1)$. The curves for the Redlich-Kwong gas with $\tilde{b}/x_1 = 0.1$ and $a/(2 b\,R\,T^{3/2}) \leq 0.1$ are not distinguished
        from their van-der-Waals counterparts.}}
    \end{center}
\end{figure}

Therefore, we now pay attention to $({\mathcal P}_m)_m$ and its
behaviors which will not depend on $r$ any longer. For a given work
output, maximizing the power output corresponds to minimizing the
cycle time $t_{cy}$ given in
(\ref{eq:total-time-of-a-single-cycle_1}). Accordingly, we can now
restrict our discussion of $({\mathcal P}_m)_m$ to the case of $\xi
\ll 1$ with $r_{12} \to \infty$ (and so $r \to \infty$). Again, we
begin with the ideal-gas case. With $r_{12} = r_{34} \to \infty$,
Eq. (\ref{eq:denominator_1}) then reduces to the simplified
expression
\begin{equation}\label{eq:den_1-1}
    \mbox{Den} = \left(\frac{1}{1 - w} + \frac{z}{w z - \beta}\,e^A\right)\,\ln r_{12}
\end{equation}
[cf. (\ref{eq:inf-thermal-reservoir-case-1-1})], and so
\begin{equation}\label{eq:power_1-1}
    {\mathcal P} = \alpha\,T_{hi}\,(1 - z)\,\left(\frac{1}{1 - w} + \frac{z}{w z - \beta}\,e^A\right)^{-1}\,.
\end{equation}
First, by requiring $\partial {\mathcal P}/\partial w
\stackrel{!}{=} 0$ and then after some steps of algebraic
manipulations, we can get
\begin{equation}\label{eq:w_result_1}
    w_m = \frac{(\beta  - z_m\,e^A) - (\beta - z_m)\,e^{A/2}}{(1 -
    e^A)\,z_m}\,.
\end{equation}
Subsequently we require $\partial {\mathcal P}/\partial z
\stackrel{!}{=} 0$ together with (\ref{eq:w_result_1}), which will
straightforwardly give rise to $z_m = \beta^{1/2}$ and so $\eta_m =
\eta_{\mbox{\tiny CAN}}$ indeed, as it is the case with $C_p \to
\infty$. Plugging this value into (\ref{eq:w_result_1}) and then
(\ref{eq:power_1-1}), it is immediate to acquire the maximum of
maximum power output given by the compression-ratio-independent
expression
\begin{equation}\label{eq:power-output-max_1}
    ({\mathcal P}_m)_m = \frac{\alpha\,T_{hi}\,(1 - \beta^{1/2})^2}{(1 + e^{A/2})^2}\,,
\end{equation}
which is one of our central results [cf. $A =
\alpha/(\dot{m}\,C_p)$]. In the limit of $C_p \to \infty$ and thus
$A \to 0^+$, Eq. (\ref{eq:w_result_1}) is confirmed to reduce to
$w_m = (\beta + z_m)/(2 z_m)$, which subsequently enables
(\ref{eq:power-output-max_1}) to reduce to the well-known
Curzon-Ahlborn-Novikov expression
\begin{equation}\label{eq:power-output-max-reservoir_1}
    {\mathcal P}_{\mbox{\tiny CAN}} = \frac{\alpha\,T_{hi}\,(1 - \beta^{1/2})^2}{4}\,,
\end{equation}
which has already been derived in many of the aforecited references,
but without considering the critical restriction of $r \to \infty$
at all.

Next, an arbitrary gas model is under consideration. Based on the
fact, explicitly derived in Sects. \ref{sec:isothermal} and
\ref{sec:iso-thermal}, that for the power output ${\mathcal P}$, the
factor $\{F(V_2, T_h^{\ast}) - F(V_1, T_h^{\ast})\}$ of the
numerator $(Q_h - Q_c)$ and the same factor of denominator $t_{cy}
\approx (t_h + t_c)$ in the limit of $\xi \ll 1$ will cancel out, it
is easy to see that Eq. (\ref{eq:power_1-1}) is valid also for an
arbitrary gas model in this limit. Therefore, the maximum of maximum
power output becomes the very expression given in
(\ref{eq:power-output-max_1}), together with $\eta_m =
\eta_{\mbox{\tiny CAN}}$, showing its universality, also within the
finite thermal reservoirs indeed, valid regardless of a specific gas
model for the working fluid (cf. \cite{ROS78} in which this scenario
has been studied for the ideal-gas model with the infinite thermal
reservoirs only, thus being simply a special case of our
discussion). It is also immediate to observe that Eq.
(\ref{eq:power-output-max_1}) reduces, in the limit of $A \ll 1$ but
with $C_p \nrightarrow \infty$, to
\begin{equation}\label{eq:power-output-max_2-1}
    ({\mathcal P}_m)_m \to \frac{\alpha\,T_{hi}\,(1 - \beta^{1/2})^2}{4}\,\left\{1 - \frac{A}{2} + {\mathcal
    O}(A)\right\}
\end{equation}
which will further reduce to (\ref{eq:power-output-max-reservoir_1})
in the limit of $\alpha \to 0$, i.e., $({\mathcal P}_m)_m \to
{\mathcal P}_m + {\mathcal O}(\alpha^2)$. The maximum of $({\mathcal
P}_m)_m$ occurs with $C_p \to \infty$, as expected.

Comments deserve here. Our finding given in
(\ref{eq:power-output-max_1}) is a generalized expression (with
physically more consistent information) of the former findings in
form of (\ref{eq:power-output-max-reservoir_1}), available in the
references (e.g., \cite{LEE90,LEE92}), which have been derived also
for the {\em finite} thermal reservoirs, without resort to any
specific gas model. From our analysis, it is then concluded that the
validity of those former findings should be limited into the regime
of both an infinitely large compression ratio $r \to \infty$ (i.e.,
approaching an infinite spatial size) and a vanishingly small heat
conductance $\alpha \to 0$ (i.e., approaching a quasi-static
process), thus corresponding to the non-realistic setup for a heat
engine! Now we should be reminded that both finite size and finite
heat-exchange rate have been introduced to overcome the drawbacks of
the Carnot cycle in original form, as discussed in Sect.
\ref{sec:introduction}. Consequently, it is claimed that the
endo-reversible approach leading to the well-known
Curzon-Ahlborn-Novikov-like results may not consistently reflect the
realistic heat engines, if applied especially to the case of finite
thermal reservoirs.

%%%%%%%%%%%%%%%%%%%%%%%%%%%%%%%%%%%%%%%%%%%%%%%%%%%%%%%%%%%%%%%%%%%%%%%
\section{Heat source and sink: Type II}\label{sec:type-2_reservoirs}
\begin{figure}[htb]
    \begin{center}
        \centering\hspace*{0cm}\vspace*{1cm}{
        \includegraphics[width=0.5\textwidth]{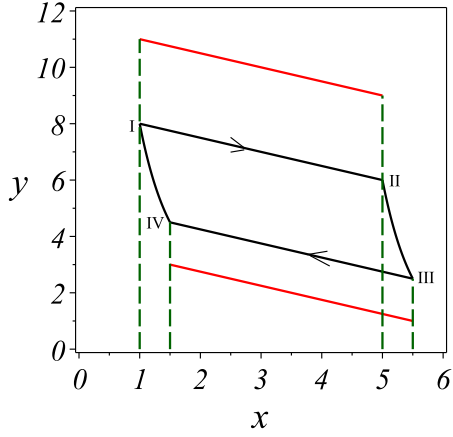}
        \caption{\label{fig:fig4}(Color online) A Lorenz cycle (I $\to$ II $\to$ III $\to$ IV $\to$
        I), denoted by the closed curve (black),
        with two finite thermal reservoirs, denoted by the two outer lines (red). The $x$-axis corresponds to the volume $V$ of a working fluid in a cylinder of engine,
        and the $y$-axis to the temperature $T$. The
        evaporator is specified by $x_1 \leq x_e \leq x_2$ with
        $x_1 = 1$ and $x_2 = 5$, while the condenser by $x_4 \leq x_c \leq x_3$ with
        $x_3 = 5.5$ and $x_4 = 1.5$. The
        temperatures of the working fluid change (``black'') as the cycle proceeds, and so
        here $T_{hi}^{\ast} = T_h(x_1) = 8$ and $T_{ho}^{\ast} = T_h(x_2) = 6$
        as well as
        $T_{ci}^{\ast} = T_c(x_3) = 2.5$ and $T_{co}^{\ast} = T_c(x_4) = 4.5$.
        The temperatures of the two reservoirs change (``red'') as the cycle exchanges
        heat, and so
        $T_{hi} = T_h(x_1) = 11$ and $T_{ho} = T_h(x_2) = 9$ as well as $T_{ci} = T_c(x_3) = 1$ and $T_{co} = T_c(x_4) = 3$.}}
    \end{center}
\end{figure}

%%%%%%%%%%%%%%%%%%%%%%%%%%%%%%%%%%%%%%%%%%%%%%%%%%%%%%%%%%%%%%%%%%%%%%%
In this type, the heat source is characterized by the {\em constant}
rate of heat input
\begin{subequations}
\begin{eqnarray}
    \dot{Q}_h &=& \alpha_h\,\{T_h(x_e) - T_h^{\ast}(x_e)\} = \alpha_h\,(T_{hi} - T_{hi}^{\ast})\label{eq:lorenz-heat-source_1}\\
    &\stackrel{!}{=}& \dot{m}_h\,C_{ph}\,(T_{hi} - T_{ho}) > 0\,,\label{eq:lorenz-heat-source_1-1}
\end{eqnarray}
\end{subequations}
in which the temperature of working fluid in the evaporator, denoted
by $T_h^{\ast}$ with $T_{hi}^{\ast} = T_h^{\ast}(x_1) \geq
T_{h}^{\ast}(x_e) \geq T_h^{\ast}(x_2) = T_{ho}^{\ast}$, also
changes in parallel with that of heating fluid (Fig.
\ref{fig:fig4}).
For a later purpose, we assume that $1 \leq T_{hi}/T_{hi}^{\ast}
\leq 2$. Combining (\ref{eq:lorenz-heat-source_1}) and
(\ref{eq:lorenz-heat-source_1-1}), we can easily express the outlet
temperature of evaporator as
\begin{equation}\label{eq:temperature-l1_1}
    T_{ho} = T_{hi} - A_h\,(T_{hi} - T_{hi}^{\ast})\,,
\end{equation}
in terms of the initially prepared temperatures $T_{hi}$ and
$T_{hi}^{\ast}$. From the condition that $T_{ho} \geq
T_{hi}^{\ast}$, it is required that $A_h \leq 1$. Similarly, the
heat sink meets the condition of heat rate given by
\begin{equation}\label{eq:lorenz-heat-sink_1}
    \dot{Q}_c = \alpha_c\,(T_{ci}^{\ast} - T_{ci}) \stackrel{!}{=} \dot{m}_c\,C_{pc}\,(T_{co} - T_{ci}) >
    0\,,
\end{equation}
which can subsequently yield
\begin{equation}\label{eq:temperature-l1_2}
    T_{co} = T_{ci} - A_c\,(T_{ci} - T_{ci}^{\ast})\,.
\end{equation}
From the condition that $T_{co} \leq T_{ci}^{\ast}$, it is also
required that $A_c \leq 1$.

We are now to complete the temperature profile of heating fluid. By
its nature this temperature should then satisfy
\begin{equation}\label{eq:temperature-gradient_1}
    \frac{d T_h}{d x_e} = -k_h\,\{T_h(x_e) - T_h^{\ast}(x_e)\} = -k_h\,(T_{hi} -
    T_{hi}^{\ast})\,.
\end{equation}
This can easily be solved to yield the linearly decreasing
temperature
\begin{equation}\label{eq:temperature-gradient_2}
    T_h(x_e) = -A_h\,(T_{hi} - T_{hi}^{\ast})\,\frac{x_e - x_1}{x_2 - x_1} +
    T_{hi}\,.
\end{equation}
It is also immediate to note that Eq.
(\ref{eq:temperature-gradient_2}) becomes equivalent to the linearly
approximated form of (\ref{eq:temperature-heat-source_1}), valid for
Type I, up to ${\mathcal O}(x_e^2)$ with $T_h^{\ast} \to
T_{hi}^{\ast}$. By construction, the temperature of working fluid in
the evaporator is then given by
\begin{equation}\label{eq:temperature-gradient_3}
    T_h^{\ast}(x_e) = -A_h\,(T_{hi} - T_{hi}^{\ast})\,\frac{x_e - x_1}{x_2 - x_1} + T_{hi}^{\ast}\,.
\end{equation}
Along the similar lines, we can get, for the temperature profile of
cooling fluid,
\begin{equation}\label{eq:temperature-gradient_1-1}
    \frac{d T_c}{d x_c} = -k_c\,\{T_c^{\ast}(x_c) - T_c(x_c)\} = -k_c\,(T_{ci}^{\ast} -
    T_{ci})\,,
\end{equation}
with the temperature of working fluid meeting the condition
$T_{ci}^{\ast} = T_c^{\ast}(x_3) \leq T_{c}^{\ast}(x_c) \leq
T_c^{\ast}(x_4) = T_{co}^{\ast}$, and so
\begin{equation}\label{eq:temperature-gradient_4}
    T_c(x_c) = -A_c\,(T_{ci}^{\ast} - T_{ci})\,\frac{x_c - x_3}{x_3 - x_4} + T_{ci}
\end{equation}
with $T_c(x_4) = T_{co} < T_c^{\ast}(x_4)$, which also becomes
equivalent to the linearly approximated form of Eq.
(\ref{eq:temperature-heat-sink_1-1}) up to ${\mathcal O}(x_c^2)$
with $T_c^{\ast} \to T_{ci}^{\ast}$. Likewise, we can determine the
temperature profile of working fluid in the condenser as
\begin{equation}\label{eq:temperature-gradient_5}
    T_c^{\ast}(x_c) = -A_c\,(T_{ci}^{\ast} - T_{ci})\,\frac{x_c - x_3}{x_3 - x_4} + T_{ci}^{\ast}\,.
\end{equation}

A comment deserves here. From Eqs.
(\ref{eq:temperature-gradient_2})-(\ref{eq:temperature-gradient_3}),
and
(\ref{eq:temperature-gradient_4})-(\ref{eq:temperature-gradient_5})
it is easy to see that if $r_{12} = x_2/x_1 \to \infty$ and so
$r_{34} = x_3/x_4 \to \infty$ as well, then $T_h(x_e) \to T_{hi}$
and $T_h^{\ast}(x_e) \to T_{hi}^{\ast}$ as well as $T_c(x_c) \to
T_{ci}$ and $T_c^{\ast}(x_c) \to T_{ci}^{\ast}$. This limiting case
exactly corresponds to the Carnot cycle in original form with
$C_{ph}, C_{pc} \to \infty$.

%%%%%%%%%%%%%%%%%%%%%%%%%%%%%%%%%%%%%%%%%%%%%%%%%%%%%%%%%%%%%%%%%%%%%%%%%%%%%%%%%
\section{Lorenz-type cycle operating in a finite time}\label{sec:lorenz}
%%%%%%%%%%%%%%%%%%%%%%%%%%%%%%%%%%%%%%%%%%%%%%%%%%%%%%%%%%%%%%%%%%%%%%%%%%%%%%%%%
In the Lorenz cycle, each isothermal process of the Carnot cycle is
replaced by the temperature-changing process (of the working fluid)
in which the ratio of its temperature change is the same as that of
the heating/sinking fluid. We begin with the ideal-gas model to
consider the rate of heat input in the evaporator
\begin{equation}\label{eq:lorenz_1}
    \dot{Q}_h = \alpha_h\,(T_{hi} - T_{hi}^{\ast}) = p\,\frac{dx_e}{dt} = \frac{nR\,T_h^{\ast}(x_e)}{x_e}\,\frac{dx_e}{dt} >
    0\,,
\end{equation}
leading to a temperature-changing expansion of the working fluid.
Substituting (\ref{eq:temperature-gradient_3}) into this, we can get
\begin{equation}\label{eq:lorenz_1-1}
    dt = \left\{-\frac{nR}{\dot{m}_h\,C_{ph}}\,\frac{x_e-x_1}{x_2-x_1} +
    \tau_h(x_1)\right\}\,\frac{dx_e}{x_e}\,,
\end{equation}
which will subsequently be integrated over $x_e$ to yield the
duration of heat input
\begin{equation}\label{eq:lorenz_2}
    t_h = \left\{\tau_h(x_1) - \frac{nR}{\dot{m}_h\,C_{ph}}\right\}\,\ln(r_{12}) +
    \frac{nR}{\dot{m}_h\,C_{ph}} \left\{\frac{r_{12}\,\ln(r_{12})}{r_{12}-1} - 1\right\} > 0
\end{equation}
with $\tau_h(x_e) = n
R\,T_{hi}^{\ast}\cdot(\alpha_h)^{-1}\,\{T_h(x_e) -
T_{hi}^{\ast}\}^{-1}$. Then the amount of heat input turns out to be
\begin{equation}\label{eq:lorenz_3}
    Q_h = \int_0^{t_h}\,\dot{Q}_h\,dt = \alpha_h\,(T_{hi} - T_{hi}^{\ast})\,t_h > 0\,.
\end{equation}

Likewise, we have in the condenser
\begin{equation}\label{eq:lorenz_4}
    \dot{Q}_c = \alpha_c\,(T_{ci} - T_{ci}^{\ast}) = p\,\frac{dx_c}{dt}
    = \frac{nR\,T_c^{\ast}(x_c)}{x_c} \frac{dx_c}{dt} < 0\,,
\end{equation}
leading to a compression of the working fluid. Substituting
(\ref{eq:temperature-gradient_5}) into this and then integrating
over $x_c$, we can get
\begin{equation}\label{eq:lorenz_5}
    t_c = \tau_c(x_3)\,\ln(r_{34}) + \frac{nR}{\dot{m}_c\,C_{pc}} \left\{\frac{r_{34}\,\ln(r_{34})}{r_{34}-1} - 1\right\} > 0
\end{equation}
with $\tau_c(x_c) = n
R\,T_{ci}^{\ast}\cdot(\alpha_c)^{-1}\,\{T_{ci}^{\ast} -
T_c(x_c)\}^{-1}$. Then, the amount of dumped heat becomes
\begin{equation}\label{eq:lorenz_6}
    |Q_c| = -\int_0^{t_c} \dot{Q}_c\,dt = \alpha_c\,(T_{ci}^{\ast} - T_{ci})\,t_c > 0\,.
\end{equation}

\begin{figure}[htb]
    \begin{center}
        \centering\hspace*{0cm}\vspace*{1cm}{
        \includegraphics[width=0.5\textwidth]{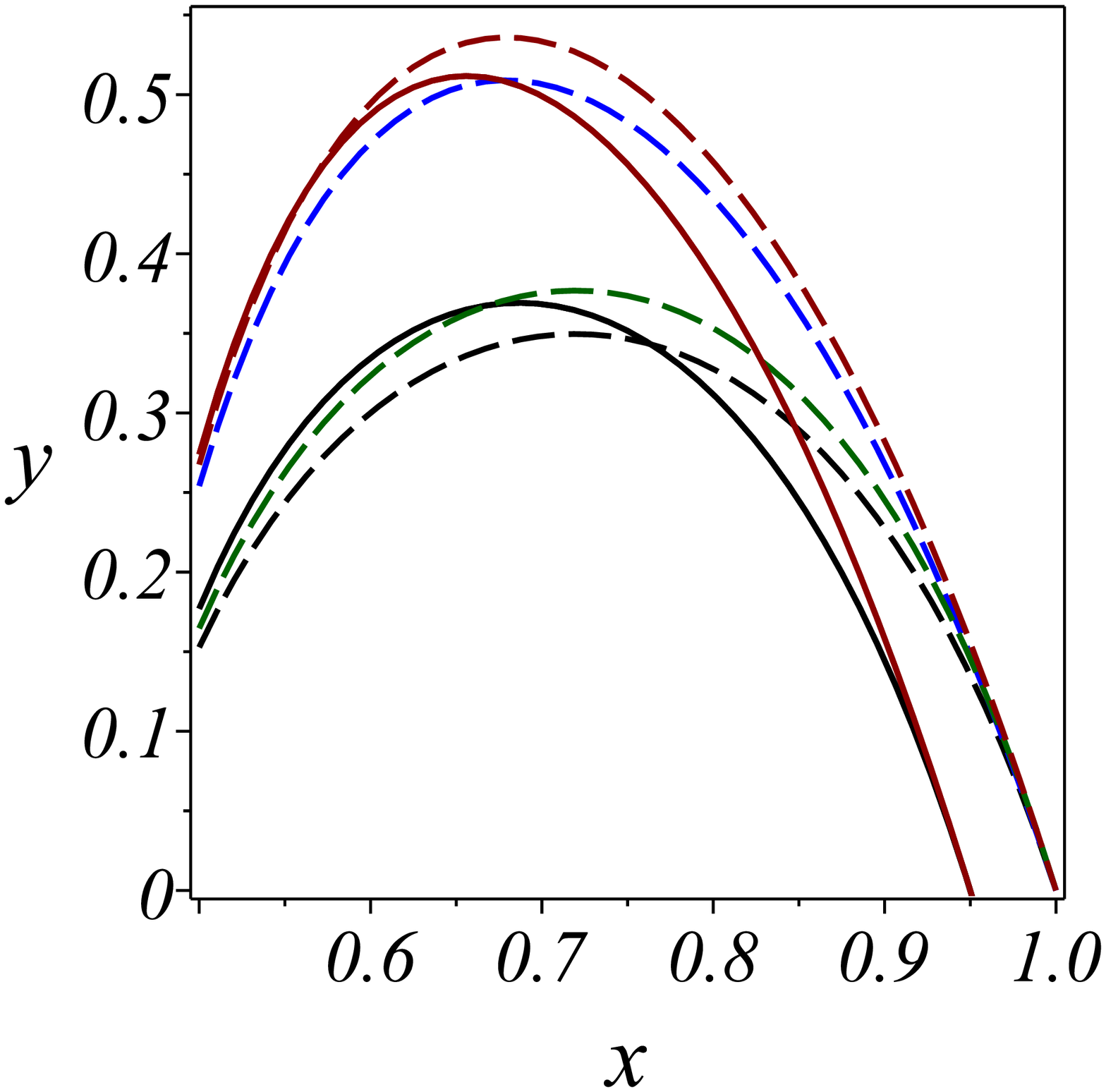}
        \caption{\label{fig:fig5}(Color online) The power output of the Lorenz cycle normalized by the Curzon-Ahlborn-Novikov
        power output,
        $y = {\mathcal P}/{\mathcal P}_{\mbox{\tiny CAN}}$ versus
        the temperature ratio,
        $x = T_{ci}^{\ast}/T_{hi}^{\ast}$, denoted by the solid
        curves, for the ideal gas (Id), the van der Waals gas (vdW) with $\tilde{b}/x_1 = 0.1$, and the Redlich-Kwong gas (RK)
        with $\tilde{b}/x_1 = 0.1$ and $a/(2 b\,R\,T^{3/2}) \leq 0.1$.
        Here the efficiency $\eta = 1 - x$. We applied Eqs. (\ref{eq:lorenz_2})-(\ref{eq:lorenz_3}) and
        (\ref{eq:lorenz_5})-(\ref{eq:lorenz_6}), here with the initial
        values, $\beta = 0.3$ and $w = 2/3$ [cf. before
        (\ref{eq:long-compression-ratio_1-2})] as well as $A = 0.1$ [cf. before
        (\ref{eq:lorenz-power-output_1})].
        In comparison, their Carnot counterparts are
        also given by the dash curves [cf. Figs. 2 and 3]. From top to bottom in form of ($\text{color}: r_{12}, x_m, y_m; \eta_m$)
        with the maximum value $y_m$ at $x = x_m$, and $\eta_m = 1 - x_m$: For the Lorenz cycle,
        (red: 5, 0.656, 0.512; 0.344); (black: 2, 0.687, 0.369; 0.313), each of which is valid for all three gas models at the same time.
        For the Carnot cycle, (red: 5, 0.679, 0.536; 0.321) for vdW and RK; (blue: 5, 0.679, 0.510; 0.321) for Id;
        (green: 2, 0.720, 0.377; 0.280) for vdW and RK; (black: 2, 0.720, 0.350; 0.280) for Id.
        The region of $x$ for $y \geq 0$ for the Lorenz curve comes from the condition that $T_{hi}^{\ast} \leq T_{ho}$ and
        $T_{ci}^{\ast} \geq T_{co}$.}}
    \end{center}
\end{figure}
Now we consider the power output ${\mathcal P}$ and its maximization
${\mathcal P}_m$ with respect to the two initially prepared
temperatures ($T_{hi}^{\ast}, T_{ci}^{\ast}$), again without
neglecting the total adiabatic time ($t_{a1} + t_{a2}$). For the
same justification as applied for the Carnot cycle, it is then
expected that the maximum value ${\mathcal P}_m$ should depend on
the compression ratio $r$. In fact, Figs. 5 and 6 show that for the
ideal gas, the van der Waals has, and the Redlich-Kwong gas, both
$({\mathcal P}_m)_{\mbox{\tiny L}}$ and $(\eta_m)_{\mbox{\tiny L}}$
of the Lorenz cycle increase with the ratio $r_{12}$ (and thus $r$),
as well as $(\eta_m)_{\mbox{\tiny L}}$ is higher than
$(\eta_m)_{\mbox{\tiny C}}$ of the Carnot cycle indeed, while
$({\mathcal P}_m)_{\mbox{\tiny C}}$ of the Carnot cycle may be
higher than $({\mathcal P}_m)_{\mbox{\tiny L}}$ when $A =
\alpha/(\dot{m}\,C_p)$ is small enough ($A = 0.1$ for Fig. 5) within
the van der Waals and the Redlich-Kwong [cf. before
(\ref{eq:lorenz-power-output_1})].
\begin{figure}[htb]
    \begin{center}
        \centering\hspace*{0cm}\vspace*{1cm}{
        \includegraphics[width=0.5\textwidth]{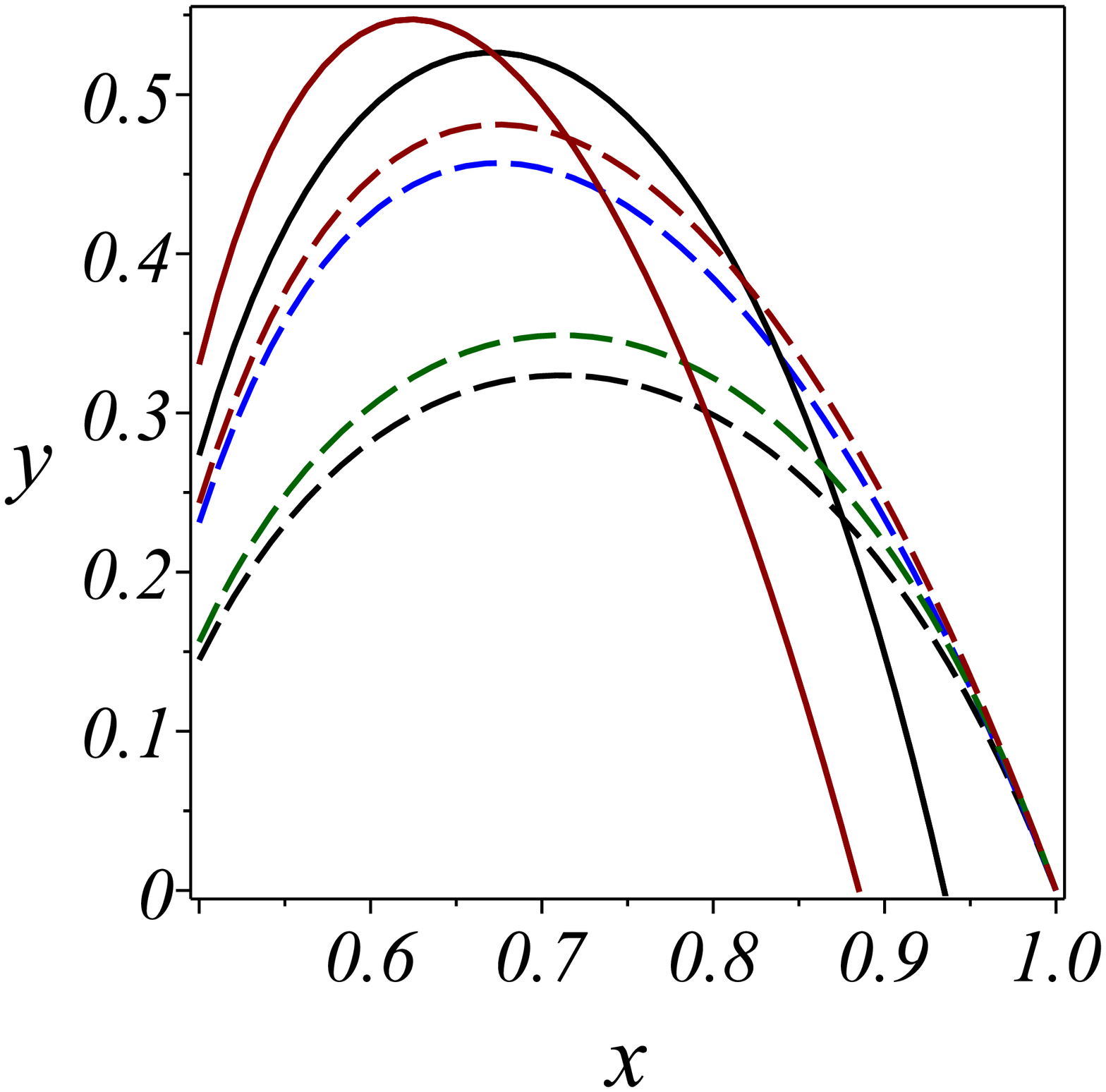}
        \caption{\label{fig:fig6}(Color online) (Color online)
        The power output of the Lorenz cycle normalized by the Curzon-Ahlborn-Novikov
        power output,
        $y = {\mathcal P}/{\mathcal P}_{\mbox{\tiny CAN}}$ versus
        the temperature ratio,
        $x = T_{ci}^{\ast}/T_{hi}^{\ast}$, given for $A = 0.4$, otherwise in the same condition as for Fig. 5.
        From top to bottom: For the Lorenz cycle,
        (red: 5, 0.642, 0.547; 0.358); (black: 2, 0.671, 0.526; 0.329),
        each of which is valid for all three gas models at the same time.
        For the Carnot cycle, (red: 5, 0.675, 0.481; 0.325) for vdW and RK; (blue: 5, 0.675, 0.457; 0.325) for Id;
        (green: 2, 0.712, 0.349; 0.288) for vdW and RK; (black: 2, 0.712, 0.324; 0.288) for Id.}}
    \end{center}
\end{figure}
Here we employ the relation given by $r_{34}/r_{12} = \{(1 -
A\,\{(w^{-1} - 1)\})(1 + A\,\{1 - \beta\,(z
w)^{-1}\})\}^{1/(\gamma-1)}$ resulting from the condition for the
two (non-negligible) adiabatic processes that
$T_{ho}^{\ast}\cdot(x_2)^{\gamma-1} =
T_{ci}^{\ast}\cdot(x_3)^{\gamma-1}$ and
$T_{hi}^{\ast}\cdot(x_1)^{\gamma-1} =
T_{co}^{\ast}\cdot(x_4)^{\gamma-1}$ [cf. before
(\ref{eq:long-compression-ratio_1-2})].

Therefore, we now restrict our discussion to the maximum of maximum
power output occurring with $r \to \infty$, again in the limit of
$\xi \ll 1$ and so $r_{12} \to \infty$, in order to compare this
with its Carnot-cycle counterpart. We again assume that $\alpha_h =
\alpha_c =: \alpha$ and $C_{ph} = C_{pc} =: C_p$ as well as
$\dot{m}_h = \dot{m}_c =: \dot{m}$, and thus $A_h = A_c =: A$. Then
we can easily obtain
\begin{equation}\label{eq:lorenz-power-output_1}
    {\mathcal P} = \frac{Q_h - |Q_c|}{t_h + t_c} =
    \frac{\alpha\,(T_{hi} - T_{hi}^{\ast})\,t_h - \alpha\,(T_{ci}^{\ast} - T_{ci}) t_c}{t_h +
    t_c}\,,
\end{equation}
where in the limit of $r_{12}=r_{34} \to \infty$ [cf.
(\ref{eq:lorenz_2}) and (\ref{eq:lorenz_5})]
\begin{eqnarray}\label{eq:long-compression-ratio_1-1}
    t_h &\to& \tau_h(x_1)\,\ln(r_{12})\\
    t_c &\to& \left\{\tau_c(x_3) +
    \frac{nR}{\dot{m}\,C_p}\right\}\,\ln(r_{12})\,.
\end{eqnarray}
In terms of the three dimensionless temperatures given by $z =
T_{ci}^{\ast}/T_{hi}^{\ast}$ and both initial values $\beta =
T_{ci}/T_{hi}$ and $w = T_{hi}^{\ast}/T_{hi}$, the power output can
then be rewritten as
\begin{equation}\label{eq:long-compression-ratio_1-2}
    {\mathcal P} = \alpha\,T_{hi}\,\{1-z-A\,(z-\beta/w)\}\,\left(\frac{1}{1-w} + \frac{z}{z w-\beta} +
    \frac{A}{w}\right)^{-1}\,.
\end{equation}

We now consider $\partial {\mathcal P}/\partial w \stackrel{!}{=} 0$
and $\partial {\mathcal P}/\partial z \stackrel{!}{=} 0$. After some
algebraic manipulations, we can then arrive at two equalities
\begin{eqnarray}
    && \frac{A\beta}{(w_m)^2}\,\left\{\frac{z_m-\beta}{(1-w_m)\,(z_m\,w_m-\beta)} +
    \frac{A}{w_m}\right\}\label{eq:long-compression-ratio_1-3-0}\\
    &=& \left\{1-z_m-A\,(z_m-\beta/w_m)\right\}\,\left\{\frac{-1}{(1-w_m)^2} + \frac{(z_m)^2}{(z_m\,w_m-\beta)^2} +
    \frac{A}{(w_m)^2}\right\}\n
\end{eqnarray}
and
\begin{equation}
    \frac{1}{1-w_m} + \frac{z_m}{z_m\,w_m-\beta} +
    \frac{A}{w_m} = \frac{1-z_m-A\,(z_m-\beta/w_m)}{1+A}\,\frac{\beta}{(z_m\,w_m-\beta)^2}\,.\label{eq:long-compression-ratio_1-3}
\end{equation}
Combining both the equalities, we can finally obtain the exact
expression
\begin{equation}\label{eq:long-compression-ratio_1-4}
    z_m = \frac{\beta}{w_m\,a}\,\frac{1 - a^2 + w_m\,(a^2 +
    a-1)}{(a+1)\,w_m-a}\,,
\end{equation}
where the dimensionless parameter $a = (A+1)^{1/2}$. Substituting
this expression into (\ref{eq:long-compression-ratio_1-3}) will
yield
\begin{equation}\label{eq:long-compression-ratio_1-5}
    w_m = \frac{a+\beta^{1/2}}{a+1}\,,
\end{equation}
and subsequently
\begin{equation}\label{eq:long-compression-ratio_1-6}
    z_m = \frac{\beta^{1/2}\,\{1+\beta^{1/2}\,(a^2+a-1)\}}{a\,(a+\beta^{1/2})} =
    \beta^{1/2}\,\left\{1-\frac{(a^2-1)(1-\beta^{1/2})}{a\,(a+\beta^{1/2})}\right\} < \beta^{1/2}\,.
\end{equation}
Then the maximum of maximum power output (in the limit of an
infinitely large compression ratio) turns out to be the
compression-ratio-independent expression
\begin{equation}\label{eq:long-compression-ratio_1-7}
    ({\mathcal P}_m)_m =
    \frac{\alpha\,T_{hi}\,(1-\beta^{1/2})^2}{(1+a)^2}\,,
\end{equation}
which is another central result. With $a \to 1$ and so $A \to 0$,
Eq. (\ref{eq:long-compression-ratio_1-7}) easily reduces to its
Curzon-Ahlborn-Novikov form given in
(\ref{eq:power-output-max-reservoir_1}), derived also for the Lorenz
cycle but again without consideration of the critical restriction of
$r \to \infty$ (e.g., \cite{LEE92}). It is straightforward to
observe that the Lorenz maximum-maximum power output $({\mathcal
P}_m)_m$ given in (\ref{eq:long-compression-ratio_1-7}) is larger
than its Carnot-cycle counterpart given in
(\ref{eq:power-output-max_1}), due to the fact that $e^A > 1+A$. In
fact, Eq. (\ref{eq:long-compression-ratio_1-7}) reduces to
(\ref{eq:power-output-max_1}) with $A \to 0$. We also see from
(\ref{eq:long-compression-ratio_1-6}) that the efficiency $\eta_m =
1 - z_m$ at the point of $({\mathcal P}_m)_m$ for the Lorenz cycle
is higher than its Carnot-cycle counterpart $\eta_{\mbox{\tiny
CAN}}$, whereas the former result available in \cite{LEE92} for the
efficiency $\eta_m$ of the Lorenz cycle has simply been shown to be
identical to that of the Carnot cycle.

Comments deserve here. Our result consists with the fact that the
Lorenz cycle may be understood as an infinite series of the Carnot
cycles operating within the finite thermal reservoirs, denoted by
$\sum_n \mbox{C}_n$ \cite{HWA05}, the efficiency of which is
actually higher than that of a single Carnot cycle $\mbox{C}_1$
\cite{OND81}. From this relation between the Lorenz and Carnot
cycles and the fact that the Carnot cycle result is valid for
arbitrary gas models of the working fluid, as rigorously discussed
in Sect. \ref{sec:carnot}, it must be true that Eq.
(\ref{eq:long-compression-ratio_1-7}) is also valid for arbitrary
gas models, but in the limit of an infinitely large compression
ratio only. In this limit ($r_{12}, r_{34} \to \infty$), the
temperature $T_h(x_e) \to T_{hi}$ in
(\ref{eq:temperature-gradient_2}) and the temperature $T_c(x_c) \to
T_{ci}$ in (\ref{eq:temperature-gradient_4}). This stands for the
infinitely large heat capacity of heating/cooling fluid ($C_p \to
\infty$) as well as the isothermal heat exchange in the evaporator
and condenser [cf. (\ref{eq:temperature-gradient_3}) and
(\ref{eq:temperature-gradient_5})], meaning that the Lorenz cycle
simply approaches a single Carnot cycle operating within the
infinite thermal reservoirs, thus being a non-realistic setup. As a
result, we may argue that the (compression-ratio-independent)
Curzon-Ahlborn-Novikov expressions, even its generalized form given
in (\ref{eq:long-compression-ratio_1-7}), should be critically
limited anyway in their own validity regimes, also for the Lorenz
cycle introduced as another model of realistic heat engine.

%%%%%%%%%%%%%%%%%%%%%%%%%%%%%%%%%%%%%%%%%%%%%%%
\section{Conclusion}\label{sec:conclusion}
In this paper, we considered the endo-reversible model of both
Carnot and Lorenz cycles operating in a finite time, and discussed
the power output and its maximization as well as the cycle
efficiency $\eta_m$ at the point of the maximized power output. To
make the cycles realistic, we employed a model of the finite heat
source and sink, and a finite duration of the adiabatic processes,
as well as the ideal-gas model followed by an arbitrary real-gas
model for the working fluid altogether. We achieved a consistent
generalization of the preceding results available in the references,
in that our results are explicitly expressed, more realistically, in
terms of not only the initially prepared temperatures of the two
finite surroundings but also their finite heat capacities and mass
flow rates as well as the heat conductance, in contrast to the
preceding results, e.g., for $\eta_m$ expressed in terms of the
initial temperatures only, and thus our resulting expressions
include the preceding ones as the special cases.

Our generalized results explicitly showed as well that the
well-known Curzon-Ahlborn-Novikov-like expressions available in many
references can be recovered, however, at most in the limit of an
infinitely large compression ratio and a vanishingly small
heat-exchange rate (even for the ideal-gas model). This limit
required as an additional condition for the validity of those
well-known expressions, albeit normally not explicitly taken into
consideration so far, gives rise to an infinitesimally slow change
in temperatures of the two surroundings, which will subsequently
lead to the limit of an infinitely large heat capacity of the
thermal reservoirs being however non-realistic again. Our finding is
certainly in contrast to the results available in many references
which have been obtained without consideration of the aforestated
critical condition. Based on our analysis with rigor, therefore, it
may be claimed that the endo-reversible model, built upon the
condition of internal reversibility, leading to those well-known
expressions is inherently limited in its consistent validity when
applied especially for heat engines interacting with the heating and
cooling fluids with a finite heat capacity and thus required to be
confined within a finite spatial size. Our concrete study also
supports the conceptual critiques on the endo-reversible approach,
developed in, e.g., \cite{GYF99,GYF02}.

Our generalized results (acquired at the macro scale) are expected
as well to contribute to providing useful guidance for a unified
understanding of multi-scale thermodynamics in a specific category,
e.g., the driven thermodynamic machines operating at the nano scale
and interacting with the finite temperature-sources in which a
conceptually more rigorous treatment of thermodynamics is often
required for the study of optimized power-output generation, because
the finite heat source and sink also are often given by certain
nano-scale forms in this domain, or even in an engineered form,
which could give rise to a considerable feedback to the machines and
so produce some highly non-trivial phenomena, possibly including a
violation of the Second Law (see, e.g., \cite{LEB16}).

\section*{Acknowledgment}
The first author appreciates the support from the Air Force Summer
Faculty Fellowship Program (USAF Contract No. FA9550-15-F-0001). He
also thanks E. Iskrenova-Ekiert and P. C. Abolmoali for their
generous hospitality during his visit to the Wright-Patterson AF
Base, as well as M. R. von Spakovsky (Virginia Tech) and J. A.
Camberos (WPAFB) for stimulating discussions. He acknowledges the
financial support provided by the Army Research Office (Grant No.
W911NF-15-1-0145).

%%%%%%%%%%%%%%%%%%%%%%%%%%%%%%%%%%%%%%%%%%%%%%%%%%%%%%%%%%%%%%%%%%%%%%%%%%%%%%%%%%%%%%%%
\appendix
\section{An arbitrary real-gas model in an adiabatic process}\label{sec:booster}
%%%%%%%%%%%%%%%%%%%%%%%%%%%%%%%%%%%%%%%%%%%%%%%%%%%%%%%%%%%%%%%%%%%%%%%%%%%%%%%%%%%%%%%%
We briefly introduce the properties of an arbitrary real-gas model
used for the discussion of Sects. \ref{sec:adiabatic} and
\ref{sec:adiabatic-2} by following the ideas developed in
\cite{TJI06}. In an adiabatic process ($\delta Q = 0$), Eqs.
(\ref{eq:1st-law_2}) and (\ref{eq:C_v-arbitrary-gas_1-0}) will give
\begin{equation}\label{eq:adiabatic-process-1-1}
    \left\{\frac{f(T)}{T} + \int^{\scriptscriptstyle V} \left(\frac{\partial^2 p}{\partial T^2}\right)_{\scriptscriptstyle V}\,dV\right\} dT
    + \left(\frac{\partial p}{\partial T}\right)_{\scriptscriptstyle V}\,dV =
    0\,.
\end{equation}
This can be rewritten as
\begin{equation}\label{eq:adiabatic-process-1-2}
    \left\{\frac{\partial}{\partial T}\,{\mathcal J}(T,V)\right\}_{\scriptscriptstyle V}\,dT +
    \left\{\frac{\partial}{\partial V}\,{\mathcal J}(T,V)\right\}_{\scriptscriptstyle T}\,dV = 0\,,
\end{equation}
where
\begin{equation}\label{eq:adiabatic-2nd-term}
    {\mathcal J}(T,V) = \int^{\scriptscriptstyle V} \left(\frac{\partial p}{\partial T}\right)_{\scriptscriptstyle V'}\,dV' +
    \int^{\scriptscriptstyle T} \frac{f(T')}{T'}\,dT
\end{equation}
with $d{\mathcal J} = 0$. Therefore, the quantity ${\mathcal
J}(T,V)$ remains unchanged in an adiabatic process. For the ideal
gas, it is easy to verify that this exactly corresponds to the
well-known invariance of $T V^{\gamma-1}$ during an adiabatic (and
reversible) process. Then we separately apply the invariant
${\mathcal J}$ for the two adiabatic processes (i.e., expansion and
compression) between $T_h^{\ast}$ and $T_c^{\ast}$, which will
immediately justify the second equality of
(\ref{eq:time-duration-2-1-adiabatic}). By noting that the second
term on the right-hand side of (\ref{eq:adiabatic-2nd-term}) is a
function of temperature only, we can also obtain
\begin{equation}\label{eq:adiabatic-invariant-1-3}
    F(V_2,T_h^{\ast}) - F(V_1,T_h^{\ast}) = F(V_3,T_c^{\ast}) -
    F(V_4,T_c^{\ast})\,,
\end{equation}
which is used for Eq. (\ref{eq:heat-arbitrary_1-0}).
\end{document}